\documentclass[12pt]{article}
\usepackage[symbol*]{footmisc}
\usepackage{array}
\usepackage{amsmath}
\usepackage{amssymb}
\usepackage{amsfonts}
\usepackage{authblk}
\usepackage[dvips]{graphicx}
\usepackage{feynmf}
\usepackage{hyperref}

\numberwithin{equation}{section}

\input epsf
\newcommand{\imm}{\mathrm{i}}
\newcommand{\esp}{\mathrm{e}}
\renewcommand{\l}{\mathcal{L}}
\newcommand{\de}{\mathrm{d}}
\newcommand{\e}{\varepsilon}
\renewcommand{\u}{\hat{u}}

\newcommand{\dslash}{\not{\hbox{\kern-2pt $\partial$}}}
\newcommand{\pslash}{\not{\hbox{\kern-2.3pt $p$}}}
 \newtoks\nslashfraction
 \nslashfraction={.13}
 \newcommand{\nslash}[1]{\setbox0\hbox{$ #1 $}
   \setbox0\hbox to \the\nslashfraction\wd0{\hss \box0}/\box0 }

\begin{document}


\begin{flushright}
{ NYU-TH-06/11/22}
\end{flushright}
\vskip 0.9cm

\centerline{\Large \bf Delocalization from Anomaly Inflow}
\centerline{\Large \bf and Intersecting Brane Dynamics}
\vspace{0.3in}
\vskip 0.7cm
\centerline{\large Luca Grisa}
\vskip 0.3cm
\centerline{\em Center for Cosmology and Particle Physics}
\centerline{\em Department of Physics, New York University, New York, NY 10003, USA}

\vskip 1.9cm

\begin{abstract}

We study intersecting D-brane models, that describe at low energies a two dimensional chiral fermion theory localized at the intersection. The fermions are coupled to gauge fields in the bulk. The resulting low energy theory is equivalent to the Gross-Neveu model with dynamical chiral symmetry breaking. No Nambu-Goldstone boson associated with spontaneously broken symmetries appears in two dimensional field theories. In the present work we discuss solvable models with the same basic dynamics of the dual Gross-Neveu model. The disappearance of the Nambu-Goldstone boson is obtained from D-brane dynamics. The mechanism relies on the non-trivial dynamics of a gauge field due to anomaly inflow.

\end{abstract}

\vspace{3cm}



\newpage

\section{Introduction}
\label{intro}
The dynamics of chiral symmetry breaking and confinement still deserves a better understanding. They are responsible of the physics of low energy composite particles, such as mesons and baryons, but quantum chromo-dynamics (QCD), the theory of strong interactions, has proven difficult to study analytically for being strongly coupled at the scale of the typical hadron mass ($\sim 1$ GeV.)

Within string theory, a strongly coupled field theory is dual to a weakly coupled gravity model, \cite{Maldacena:1997re}\nocite{Gubser:1998bc}-\cite{Witten:1998qj}. This opens a possibility to infer physics of strongly coupled field theories from the semi-classical dynamics in their dual weakly coupled picture.

Recently
\cite{Sakai:2004cn}\nocite{Babington:2003vm}\nocite{Evans:2004ia}\nocite{Kruczenski:2003uq}\nocite{Karch:2002sh}\nocite{Karch:2002xe}\nocite{Kruczenski:2003be}\nocite{Maldacena:2000yy}\nocite{Casero:2005se}-\cite{Erdmenger:2004dk},
this idea has been thoroughly investigated using brane configurations. These constructions have been used as tools to study low energy dynamics and often to compare with results of QCD-like models. Chiral symmetry breaking and confinement are well described therein and the spectrum resembles the bound state spectrum found in low energy QCD.

\begin{figure}[t]
\begin{center}
\includegraphics[width=0.5\textwidth]{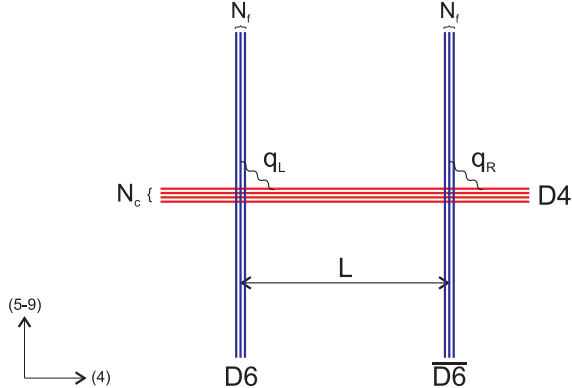}
\caption{\small\emph{Configuration of $N_c$ $D4$-branes and $N_f$ $D6$- and $\overline{D6}$-branes.}}
\label{straight}
\end{center}
\end{figure}

In the present work, we discuss chiral symmetry breaking in two dimensional field theories and the dynamics that leads to the disappearance of the Nambu-Goldstone~boson in the dual gravity picture. Let us consider the model proposed in \cite{Antonyan:2006qy}. The configuration consists of $N_c$ $D4$-branes intersecting with $N_f$ $D6$- and $\overline{D6}$-branes, see Fig.~\ref{straight}.
The intersection is chosen to be two dimensional. At low energy, this brane configuration reduces to a two dimensional field theory of chiral fermions interacting with a bulk $U(N_c)$ gauge field. The fermion fields are in the fundamental representation of the chiral group $U(N_f)\times U(N_f)$. After integrating the gauge field out, an effective four-fermion interaction is generated. The low energy theory is therefore equivalent to the (generalized) Gross-Neveu (GN) model \cite{Gross:1974jv}.

In the GN model, chiral symmetry is dynamically broken for any values of the parameters. From general grounds, the Nambu-Goldstone (NG) boson associated with chiral symmetry breaking is described by the massless Kaluza-Klein (KK) mode of the gauge field propagating on the $D6$-brane worldvolume. Its absence is required by Coleman's theorem \cite{Coleman:1973ci}. In \cite{Antonyan:2006qy}, a mechanism is suggested to remove the zero-mode from the low energy spectrum.

The proposal is related to the dynamics described in \cite{Itzhaki:2005tu} to solve a paradox, that seems to appear in brane configurations with two dimensional intersection. It seems na\"{\i}vely that the physics of the weakly coupled theory and its strongly coupled dual are very different. The former suggests the theory to be invariant under eight supercharges and consisting of chiral fermions localized on the intersection. On the other hand, the dual strongly coupled description leads to a rather different picture, where the theory is invariant under three dimensional Poincar\'e group, supersymmetry is enhanced to sixteen supercharges and a mass gap is present in the spectrum. The paradox is solved by the non-trivial dynamics induced by anomaly inflow: the non-zero chiral anomaly on the two dimensional intersection is cancelled by the anomalous gauge transformation of the bulk action. As it should become clear later, the anomaly inflow changes the dynamics of the gauge field to the extent of delocalizing it away from the intersection. In the proximity of the intersection no modes are present and the physics is described by a Chern-Simon gauge theory on three dimensional space-time.

We will show that in the dual GN model the delocalization of the gauge field explains the absence of the NG boson in the low energy spectrum. The dynamics can be referred to as the dual picture of the Coleman's theorem.

For simplicity, we will address the discussion of the dynamics described before in solvable $T$-dual models with abelian groups. Because the results we will present depend on the dynamics localized on the intersection, we conjecture them to hold for the non-abelian case as well.

The brane configurations consist of orthogonal $D$-branes intersecting on a two dimensional worldsheet. Specifically, we consider one $D2$-brane intersecting $D8$-branes, and one $D3$-brane intersecting $D7$-branes. Chiral fermions are localized on the intersection and coupled to the gauge fields propagating on the brane worldvolumes. One of the two gauge fields is decoupled from the low energy spectrum, so only one is dynamical. The low energy physics can be described by chiral fermions localized on a string. The fermions are interacting with an abelian gauge field propagating in the bulk.

Since the fermions have a definite chirality, their coupling with the gauge field is anomalous. Consistency of the model requires the anomaly to be cancelled. For chiral fermions localized on two dimensional topological defects, the dynamics of the gauge field in the bulk yields to the anomaly cancellation. Quantum corrections induce a topological Chern-Simon term for the gauge field in the bulk. The gauge variation of this term cancels exactly the chiral anomaly. This mechanism is known as anomaly inflow, see \cite{Callan:1984sa} and \cite{Green:1996dd}.

We will solve the equation of motion for the gauge field. The effect of the effective Chern-Simon term is to create a repulsive potential for the equivalent Schr\"odinger problem. The result is that the KK zero-mode of the gauge field has no support on the intersection and is therefore absent from the low energy spectrum of the two dimensional theory.

We argue that, in the dual GN model, the delocalization of the gauge field removes the NG boson from the low energy spectrum. As suggested in \cite{Antonyan:2006qy}, the reason is that the zero-mode of the gauge field describes the NG boson in the low energy theory. Its vanishing wavefunction at the intersection thus implies the absence of the NG boson from the two dimensional theory, where chiral symmetry is dynamically broken.

More specifically, we firstly consider the model containing a $D2$-brane intersecting $D8$-branes. On the two dimensional intersection, chiral fermions are localized and naturally coupled to the gauge field propagating on the $D2$-brane. The presence of chiral fermions yields to a non-zero anomaly. Fermion one-loop correction generates an effective Chern-Simon term for the gauge field. Its gauge variation is localized on the intersection and opposite to the chiral anomaly. Thus taking into account both effects, the action is gauge invariant. This topological term, essential for the consistency of the theory, changes the dynamics of the gauge field and - as we will see - forces the field to delocalize in the bulk. Therefore the zero-mode of the fluctuations of the field is absent in the two dimensional theory.

We then consider a $D3$-brane intersecting $D7$-branes. As before, the spectrum consists of chiral fermions and gauge fields. The intersection is magnetically charged under the Ramond-Ramond (RR) scalar of type IIB superstring theory. The RR field and the gauge field are coupled, and by integrating out the scalar - its flux fixed by the charge on the intersection - an effective Chern-Simon term appears for the gauge field. This term cancels the chiral anomaly and, as it will be discussed, it delocalizes the gauge field in the bulk, effectively removing the zero-mode from the low energy spectrum.

Despite some differences in the two models, the basic dynamics of the gauge field is the same: an effective topological term is generated in the bulk and forces the gauge field away from the intersection. In the case of the dual Gross-Neveu model, the effect is to remove the NG boson associated with the chiral symmetry breaking. The NG boson is described by the zero-mode of the fluctuations of the gauge field on the $D6$-brane worldvolume. The dynamics induced by the effective Chern-Simon term suppresses the zero-mode on the intersection and so it is absent in the two dimensional spectrum. The dynamics briefly portrayed can be thought as the dual picture of the Coleman's theorem valid for two dimensional field theories.

For sake of completeness, we will give a more detailed review of the dual Gross-Neveu model \cite{Antonyan:2006qy}, which is the main motivation for the present work.

Let us consider two dimensional theory of chiral fermions, non-abelian $U(N_c)$ gauge field and a global $U(N_f)\times U(N_f)$ chiral symmetry. The dual brane configuration consists of two parallel stacks of $N_f$ $D6$- and $N_f$ $\overline{D6}$-brane fixed in their orthogonal direction at a distance $L$. They both intersect $N_c$ coincident $D4$-brane on a two dimensional worldsheet. The configuration is shown in Fig.~\ref{straight}.

We can consider the 't Hooft limit $N_c\rightarrow\infty$, $g_s\rightarrow0$ and $g_sN_c$ and $N_f$ held fixed. In this limit the coupling of the strings stretching between the $D6$-branes\footnote{If not noted otherwise, we will not distinguish between $D6$-brane and $\overline{D6}$-brane.} goes to zero and their excitations become non-dynamical. The fermions in the adjont representation of $U(N_c)$ on the $D4$-brane worldvolume are decoupled from the low energy spectrum as they acquire mass via Scherk-Schwarz mechanism.

The spectrum (ignoring scalar fields) is composed of modes localized on the intersection and modes propagating in the bulk, that is on the $D4$-branes. The former are massless excitations of open strings stretching between $D4$- and $D6$-branes and are chiral fermions in the fundamental representation of both the chiral and the gauge group. One of the chirality is removed by the GSO projection and, since the GSO projection acting on $4-\bar6$ strings is opposite to that on $4-6$ strings, the chirality is different for different intersections. Thus, the spectrum contains both left and right chiral fermions; being localized on different intersections, they interact only through the exchange of the bulk $U(N_c)$ gauge field, which arises from the strings ending on the $D4$-branes. At low energy, we can integrate the gauge field out yielding an effective non-local four-fermion interaction. The two dimensional effective theory is therefore equivalent to the (generalized) Gross-Neveu (GN) model \cite{Gross:1974jv}.

Chiral symmetry breaking has been discussed for the dual GN model in \cite{Antonyan:2006qy}. It is described therein as the dynamics of $N_f$ $D6$-branes on the background generated by $N_c$ $D4$-branes, when $N_c$ is much larger than $N_f$. The equation of motion for the $D6$-branes has two different solutions.

\begin{figure}[t]
\begin{center}
\includegraphics[width=0.5\textwidth]{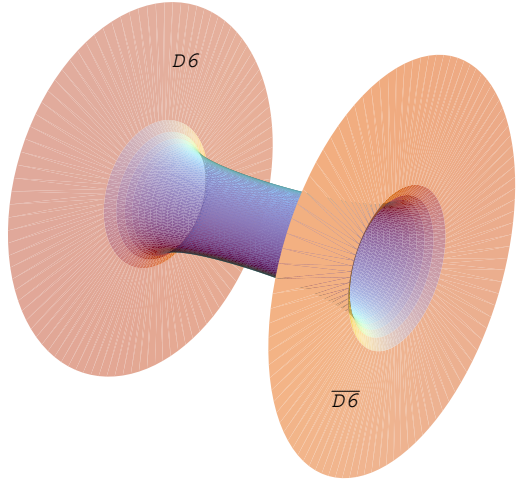}
\caption{\small\emph{The $D6$- and $\overline{D6}$-brane are connected through a wormhole.}}
\label{worm}
\end{center}
\end{figure}

In the first solution, $D6$- and $\overline{D6}$-branes are distinct and separated by a distance $L$ in the transverse direction. This solution preserves the chiral symmetry $U(N_f)\times U(N_f)$. In the second solution, $D6$- and $\overline{D6}$-branes join together through a wormhole (see Fig.~\ref{worm}.) The fact that $D6$- and $\overline{D6}$-brane are a single connected brane implies that the chiral symmetry group is broken down to the diagonal $U(N_f)$.

In the case of zero temperature the wormhole configuration has lower energy than the other, thus suggesting that the chiral symmetry is spontaneously broken for any value of the parameters, as found in \cite{Gross:1974jv}. Above a critical temperature chiral symmetry is restored, as the system undergoes a geometrical transition from the configuration with the wormhole to the one with two distinct sets of $D6$-branes. Analogously, in \cite{Aharony:2006da} a confinement-deconfinement phase transition is discussed.\\

The paper is organized as follows. In section~\ref{codim1} we consider the $D2$- and $D8$-brane configuration, and in section~\ref{codim2} the one with $D3$- and $D7$-branes. We then briefly discuss some aspects of the dynamics for more general models in section~\ref{codimgen} and conclude in section~\ref{disc}. In the appendices~\ref{app1} and \ref{app2} we summarize the known facts about the bosonization rules in the Schwinger model for Dirac fermions and chiral fermions respectively. The appendix~\ref{app3} explains certain technical aspects we encounter in section~\ref{codim2}.

\section{Localized Chiral Fermion in $\mathbf{\mathbb{R}^{2,1}}$}
\label{codim1}

As briefly presented in the introduction, the model we would like to discuss is the low energy effective theory of the brane configuration, composed of one $D2$-brane orthogonal to $D8$-branes. The intersection is chosen to be two dimensional and it is topologically equivalent to a (long) string.\\
The low energy spectrum of the model is found by looking at the massless excitations of the open strings stretching between the branes. They can be distinguished in $8-8$ strings ending on the $D8$-brane, $2-2$ strings ending on the $D2$-brane, and $2-8$ strings stretching between $D2$-brane and $D8$-brane.\\
We consider the simplified case where excitations of $8-8$ strings are decoupled from the low energy physics. The low energy excitations of $2-2$ strings are, while neglecting scalar modes, a gauge field $A_M$ and a fermion transforming in the adjoint representation of the gauge group. The fermion is massive and decoupled from the low energy spectrum by non-supersymmetrically compactifying a direction along the $D2$-brane.\\
The open string stretching between the $D2$- and $D8$-brane gives massless fermions $\psi_\mathrm{L}$ localized on the intersection. One of the chirality is removed by the GSO projection, therefore the fermion is chiral. The corresponding low-energy effective Lagrangian is
\begin{equation}
	\l=-\frac{1}{4g^2}F_{MN}^{\,2}+
	\delta(u)\,\bar\psi_\mathrm{L}(\imm\nslash\partial+\nslash A)\psi_\mathrm{L}\,,
\end{equation}
the chiral fermion is naturally coupled to the gauge field.\\
On top of this classical Lagrangian, we have to consider the term that is induced by radiative corrections; due to a fermion one-loop diagram, the following current
\begin{equation}
\langle J^M\rangle\sim\epsilon^{MNP}\theta(u)F_{NP}
\end{equation}
is generated in the bulk. Including this effect, the effective Lagrangian reads
\begin{equation}
	\l=-\frac{1}{4g^2}F_{MN}^{\,2}-\frac{1}{2}\theta(u)\epsilon^{MNP}A_MF_{NP}+
	\delta(u)\,\bar\psi_\mathrm{L}(\imm\nslash\partial+\nslash A)\psi_\mathrm{L}\,.
\end{equation}
In two dimensions a well-known fact is the dual relation between fermions and boson. Using the bosonization rules discussed in the appendices \ref{app2} for chiral fermions coupled to a gauge field, the dual Lagrangian
\begin{eqnarray}
	\l=&-&\frac{1}{4g^2}F_{MN}^{\,2}-\frac{1}{2}\theta(u)\epsilon^{MNP}A_MF_{NP}+\label{eff.lagr}\nonumber\\
	   &+&\delta(u)\left[
	\frac{1}{2}(\partial_\mu\,\phi)^2+A_\mu(\eta^{\mu\nu}-\epsilon^{\mu\nu})\partial_\mu\phi+\frac{1}{2}A_\mu^2
	\right]\label{lagr}\,.
\end{eqnarray}

The presence of chiral fermions leads to a non-zero anomaly. The fermion $\psi_\mathrm{L}$ is proportional to $\exp[\imm\phi]$, thus the gauge transformation $\delta\psi_\mathrm{L}=\exp[-\imm\Lambda]$ translates into a shift for the dual boson $\delta\phi=-\Lambda$. Under this transformation, the term localized on the intersection is not invariant:
\begin{equation}
\label{anomaly}
\delta\left[
	\frac{1}{2}(\partial_\mu\,\phi)^2+A_\mu(\eta^{\mu\nu}-\epsilon^{\mu\nu})\partial_\mu\phi+\frac{1}{2}A_\mu^2\right]\delta(u)
	=\Lambda\,(\epsilon^{\mu\nu}\partial_\mu A_\nu)\,\delta(u)\,.
\end{equation}
Under gauge transformation, the vector field $A_M$ transform as $\delta A_M=\partial_M\Lambda$. Because of the Chern-Simon term, the action is not gauge invariant. In fact its variation is proportional to a term localized on the intersection. This localized term - as it can be seen directly - is equal and opposite to the anomaly \eqref{anomaly}:
\begin{equation}
\delta\left[-\frac{1}{4g^2}F_{MN}^{\,2}-\frac{1}{2}\theta(u)\epsilon^{MNP}A_MF_{NP}\right]
	=-\Lambda\,(\epsilon^{\mu\nu}\partial_\mu A_\nu)\,\delta(u)\,.
\end{equation}
Hence the action is gauge invariant as a whole.

Both the bulk and the localized action are not gauge invariant, the former because of a boundary term,
the latter because of gauge anomaly. The two gauge variations are opposite of one another, therefore the anomaly is cancelled by the dynamics in the bulk. This mechanism, known as anomaly inflow, was discovered in \cite{Callan:1984sa} for topological defects in field theory and subsequently generalized for the case of intersecting branes in \cite{Green:1996dd}.

\subsection{St\"uckelberg mechanism and anomaly inflow}
We can choose a combination of the gauge and scalar fields, $B_M=A_M-\partial_M\phi$, such to diagonalize the lagrangian. The localized term reads in the new fields
\begin{equation}
	\label{prev.expr}
	\frac{1}{2}B_\mu^2-\epsilon^{\mu\nu}\partial_\mu\phi B_\nu\,.
\end{equation}
The last term on the right hand side of \eqref{prev.expr} is related to the chiral anomaly, this can be seen by integrating by
parts and expressing the term as $\phi\,\epsilon^{\mu\nu}\partial_\mu B_\nu$, see \eqref{anomaly}. This is to be cancelled by the inflow current from the bulk Chern-Simon term.\\
The bulk Lagrangian becomes in term of $B_M$
\begin{equation}
	-\frac{1}{4g^2}F_{MN}^{\,2}-\frac{1}{2}\theta(u)\epsilon^{MNP}B_MF_{NP}-
	\theta(u)\epsilon^{MNP}\partial_M\phi\,\partial_NB_P\,.
\end{equation}
The last term is a total derivative as it can be seen by integrating by parts:
\begin{equation}
	-\theta(u)\epsilon^{MNP}\partial_M\phi\,\partial_NB_P\rightarrow
	-\epsilon^{MNP}\partial_M\left[\theta(u)\phi\,\partial_NB_P\right]\rightarrow
	-\delta(u)\epsilon^{\mu\nu}\phi\,\partial_\mu B_\nu\label{inflow}\,.
\end{equation}
In the last step the Stokes' theorem has been applied. One has to take care of the presence of the step
function which ultimately leads to the localized term, roughly because
$\partial_u\,\theta(u)=\delta(u)$.\footnote{the step function is defined as +1/2 for $u>0$ and -1/2 for $u<0$.}. The localized term cancels the anomaly as noted before.\\
The action\footnote{A note should be added on the dimension of the fields: in a 1+1 worldvolume a boson field
has zero canonical dimension and a fermion 1/2; in 2+1 a boson 1/2, a fermion 1. Being the vector field
propagating in the bulk we are implicitly assuming it to have dimension 1/2 and therefore the charge of the
fermion localized on the string has dimension 1/2 as well; by substituting $A_\mu$ with $B_\mu$ we are
neglecting this charge, and implicitly redefining $B_\mu=eA_\mu-\partial_\mu\phi$ (dimension 1)
and $g^2\mapsto g^2e^2$ (dimension 1.)} reads
\begin{equation}
S=\int\de^3x\left[
	-\frac{1}{4g^2}F_{MN}^{\,2}-\frac{1}{2}\theta(u)\epsilon^{MNP}B_MF_{NP}+\delta(u)\frac{1}{2}B_\mu^2\right].
\end{equation}
In the bulk the model is described by a $U(1)$ Chern-Simon gauge theory. The topological term, as it is known \cite{Deser:1982vy}, will generate a mass gap in the bulk. The interaction of chiral fermions and gauge field induces an effective localized mass term, whose gauge variation is cancelled by the variation of the Chern-Simon term.

\subsection{A delocalized gauge field}
\label{sol}
In the present section we solve the equation of motion for $F_{MN}$. It is easier to
use lightcone coordinates defined as $x^\pm\equiv x^0\pm x^1$ on the string worldsheet,
and $u$ the coordinate transverse to the string. The Lagrangian is rewritten, using
$\eta^{+-}=\eta^{uu}=1$, $\eta^{++}=\eta^{--}=\eta^{u\pm}=0$, as (we replace $B_M$ with $A_M$)
\begin{equation}
\l=\frac{1}{2g^2}F_{+-}^{\,2}-\frac{1}{g^2}F_{u+}F_{u-}-
	\frac{1}{2}\theta(u)(A_+F_{-u}+A_-F_{u+}+A_uF_{+-})+\delta(u)A_+A_-\,.
\end{equation}
Taking variation of the previous Lagrangian with respect to $A_+$, $A_-$ and $A_u$, the equations of motion
are respectively
\begin{eqnarray}
	 \theta(u)F_{u-}+\frac{1}{g^2}(\partial_-F_{+-}+\partial_uF_{u-})
			+\delta(u)\,\frac{3}{2}A_-=0\label{eom1}\,,\\
	-\theta(u)F_{u+}+\frac{1}{g^2}(\partial_+F_{-+}+\partial_uF_{u+})
			+\delta(u)\,\frac{1}{2}A_+=0\label{eom2}\,,\\
	 \theta(u)F_{+-}+\frac{1}{g^2}(\partial_+F_{u-}+\partial_-F_{u+})=0\label{eom3}\,.
\end{eqnarray}
After algebraic manipulation\footnote{Add and subtract the differential of \eqref{eom1} with \eqref{eom2}
with respect to $x^\pm$ and use \eqref{eom3}.} we can write a differential equation for the $(+-)$-component
of $F_{MN}$ alone.\\
The technique we are using is somewhat non-standard since we solve for $F_{MN}$ instead of solving for the vector potential $A_{M}$; here we should keep in mind that the vector field $A$ must be continuous across the boundary but not its derivative along $u$, so some components of the field strength $F_{MN}$ would not be continuous and their discontinuity is determined by the boundary conditions.\\
The equation to solve for $F_{+-}$ is:
\begin{equation}
	\partial_u^{\,2}F_{+-}+2\partial_+\partial_-F_{+-}-g^4F_{+-}=0\label{eqF+-}\,.
\end{equation}
We are interested in the Kaluza-Klein spectrum for the gauge field on the string worldsheet.
Using the ansatz $2\partial_+\partial_-F_{+-}=\mu^2F_{+-}$ for massive gauge field in two dimension, the
equation for $F_{+-}$ becomes
\begin{equation}
	\partial_u^{\,2}F_{+-}-\frac{1}{4}\hat{g}^4F_{+-}=0
	\qquad\mathrm{with}\ \ \hat{g}^4\equiv g^4-4\mu^2.
\end{equation}
\paragraph{Localized Modes: $\mathbf{0\le\mu^2<g^4/4}$.}
We turn our attention to localized massive modes. In order for $A_\mu$ to be localized its mass has
to be less than $g^4/4$. This can be easily understood physically: in the bulk the presence of the
Chern-Simon term generates a mass gap between the vector field modes localized on the string and the ones in the bulk, so, when their energy is high enough, they are free to propagate in the bulk, otherwise they are constrained on the intersection.\\
The generic solution for $F_{+-}$ is
\begin{equation}
	F_{+-}=f(x^\pm)\,\exp{[-\frac{\hat{g}^2}{2}|u|]}\,,
\end{equation}
where we have imposed continuity condition on $F_{+-}$, since its definition does not contain any derivative
of the field in respect to $u$, and normalizability along the transverse direction.
The function $f(x^\pm)$ is the solution for $2\partial_+\partial_-f=\mu^2f$:
\begin{equation}
	f(x^\pm)=a\,\sin[k\cdot x]+b\,\cos[k\cdot x]\,,
\end{equation}
where $k_+k_-=-\mu^2/2$ and $k\cdot x=k_+x^++k_-x^-=k_+x_-+k_-x_+$. We solve \eqref{eom1} and
\eqref{eom2} using $F_{+-}$ given above, and find the general solution in the bulk to be:
\begin{eqnarray}
	F_{+-}&=&\phantom{-}f(x^\pm)\,\exp{[-\frac{\hat{g}}{2}^2|u|]}\,,\\
	F_{u+}&=&-\theta(u)\frac{4\partial_+f}{\hat{g}^2+g^2}\exp{[-\frac{\hat{g}}{2}^2|u|]}\,,\\
	F_{u-}&=&\phantom{-}\theta(u)\frac{4\partial_-f}{\hat{g}^2-g^2}\exp{[-\frac{\hat{g}}{2}|u|]}\,.
\end{eqnarray}
We can check that \eqref{eom3} is trivially satisfied by the solution, given $2\partial_+\partial_-f=\mu^2f$.

One more constraint must be imposed on the solution, namely the boundary condition arising from the
presence of localized fermions on the string, or alternatively the localized mass for $A_\mu$.
The boundary conditions we have to impose are continuity of the vector field $A_M$ across the border, but not
of its derivative along the transverse direction $u$. The variation of $F_{u\pm}$ is determined by
integrating the equations of motion across the string:
\begin{eqnarray}
	\phantom.[\partial_+F_{u-}]-[\partial_-F_{u+}]&=&-g^2F_{+-}\,,\label{bc.minus}\\
	\phantom.[\partial_+F_{u-}]+[\partial_-F_{u+}]&=&-\frac{1}{2}g^2F_{+-}\,,\label{bc.plus}
\end{eqnarray}
where $[F_{u\pm}]\equiv\lim_{u\to0^+}F_{u\pm}-\lim_{u\to0^-}F_{u\pm}$. Using the general solution found above,
we have the following equations to satisfy for the coefficient function $f(x^\pm)$
\begin{eqnarray}
	\hat{g}^2f&=&g^2f\,,\\
	g^2f&=&\frac{1}{2}g^2f\,.
\end{eqnarray}
Only the trivial solution satisfies this set of equations and therefore no localized modes are present
in the spectrum. The Chern-Simon term, as we have seen, generates a mass gap between the brane and the bulk
and one may have hoped to be able to localize the gauge field on the string. This is indeed not the case, the reason
is understood by looking closely at \eqref{eom1} and \eqref{eom2} and noticing that the effective potential
generated by the chiral fermions, $V_\mathrm{eff}\sim+\delta(u)A_\pm$, in the equivalent Schr\"odinger equation 
is always repulsive and therefore it can not support a localized solution.\\
We may need to emphasize that this result is valid for massive modes as well as for the zero-mode mode, so no
massless mode is present in the spectrum.
\paragraph{Continuous Modes: $\mathbf{\mu^2\ge g^4/4}$.}
We now consider the continuous modes with masses greater than $g^4/4$. The generic solution,
$k^2$ is defined as $k^4=4\mu^2-g^4$, is for $u>0$
\begin{eqnarray}
	F_{+-}&=&f_1\cos\frac{k^2u}{2}+f_2^{(+)}\sin\frac{k^2u}{2}\,,\\
	F_{u+}&=&\frac{-g^2\partial_+f_1-k^2\partial_+f_2^{(+)}}{2\mu^2}\cos\frac{k^2u}{2}+
		 \frac{ k^2\partial_+f_1-g^2\partial_+f_2^{(+)}}{2\mu^2}\sin\frac{k^2u}{2}\,,\\
	F_{u-}&=&\frac{-g^2\partial_-f_1+k^2\partial_-f_2^{(+)}}{2\mu^2}\cos\frac{k^2u}{2}-
		 \frac{ k^2\partial_-f_1+g^2\partial_-f_2^{(+)}}{2\mu^2}\sin\frac{k^2u}{2}\,,
\end{eqnarray}
and for $u<0$
\begin{eqnarray}
	F_{+-}&=&f_1\cos\frac{k^2u}{2}+f_2^{(-)}\sin\frac{k^2u}{2}\,,\\
	F_{u+}&=&\frac{ g^2\partial_+f_1-k^2\partial_+f_2^{(-)}}{2\mu^2}\cos\frac{k^2u}{2}+
		 \frac{ k^2\partial_+f_1+g^2\partial_+f_2^{(-)}}{2\mu^2}\sin\frac{k^2u}{2}\,,\\
	F_{u-}&=&\frac{ g^2\partial_-f_1+k^2\partial_-f_2^{(-)}}{2\mu^2}\cos\frac{k^2u}{2}-
		 \frac{ k^2\partial_-f_1-g^2\partial_-f_2^{(-)}}{2\mu^2}\sin\frac{k^2u}{2}\,.
\end{eqnarray}
The coefficients $f_i$ are solutions for $2\partial_+\partial_-f_i=\mu^2f_i$ and they will be fixed by the boundary
conditions. By using the previous explicit solution, the variations of $F_{u\pm}$ across the boundary are found to be:
\begin{eqnarray}
	\phantom.[\partial_+F_{u-}]-[\partial_-F_{u+}]&=&\frac{1}{2}k^2(f_2^{(+)}-f_2^{(-)})\,,\\
	\phantom.[\partial_+F_{u-}]+[\partial_-F_{u+}]&=&-g^2f_1\,.
\end{eqnarray}
The boundary conditions after substituting them into \eqref{bc.minus} and \eqref{bc.plus} are 
\begin{eqnarray}
	\frac{1}{2}k^2(f_2^{(+)}-f_2^{(-)})&=&-g^2f_1\,,\\
	g^2f_1&=&\frac{1}{2}g^2f_1\,.
\end{eqnarray}
The only acceptable solution for $f_1$ is the trivial one. With $f_1=0$ we find that the coefficients $f_2^{(\pm)}$
are equal. The generic solution for $F_{MN}$ with $\mu^2\ge g^4/4$ is shown below
\begin{eqnarray}
	F_{+-}&=&\phantom{-}f(x^\pm)\sin\frac{k^2u}{2}\,,\\
	F_{u+}&=&-\frac{\partial_+f}{2\mu^2}\left[
		 k^2\cos\frac{k^2u}{2}+2\theta(u)g^2\sin\frac{k^2u}{2}\right]\,,\\
	F_{u-}&=&\phantom{-}\frac{\partial_-f}{2\mu^2}\left[
		 k^2\cos\frac{k^2u}{2}-2\theta(u)g^2\sin\frac{k^2u}{2}\right]\,.
\end{eqnarray}
We notice that while $F_{+-}$ is continuous, $F_{u\pm}$ shifts of a phase equal to $2\arctan[k^2/g^2]+\pi$, as it crosses the string.
What may surprise is that no field is present on the worldsheet since $F_{+-}$ is identically
zero for $u\rightarrow0$. Thus, on the string, only a transverse electromagnetic field is present.
An argument for understanding this effect comes from an analogy with superconductors as ultimately the string
is in a superconducting (or Higgs) phase \cite{Witten:1984eb}: the Meissner effect is the well-known phenomenon for which
a superconductor expels the electromagnetic field from its interior. Here the superconducting string
does not allow the field $F_{+-}$ to be non-zero on its worldsheet, namely the field strength
is perfectly repelled by the superconductor. The gauge field delocalizes in the bulk and a mass gap $\sim g^4$ is generated.

\paragraph{Solution with $\mathbf{F_{+-}=0}$.}
We are now looking for a particular kind of solution found in \cite{Itzhaki:2005tu}. Those are solutions
for which we assume $F_{+-}$ to be identically zero everywhere, both on the string and in the bulk. The
equations of motion are greatly simplified
\begin{eqnarray}
	\partial^2_u\,A_-+g^2\theta(u)\partial_u\,A_-+\delta(u)\,\frac{3}{2}A_-&=&0\,,\\
	\partial^2_u\,A_+-g^2\theta(u)\partial_u\,A_++\delta(u)\,\frac{1}{2}A_+&=&0\,,\\
	\partial_u(\partial_+A_-+\partial_-A_+)&=&0\,,
\end{eqnarray}
where we have chosen the gauge $A_u=0$. This choice does not fix the gauge completely. Indeed a gauge transformation
with parameter depending only on $x^\pm$ preserves it. One may also want to fix this
residual gauge freedom, but it is unessential in the present discussion.\\
The third equation, along with $F_{+-}=0$, imposes the vector fields $A_\pm$ to be holomorphic in $x^\pm$ respectively. The solution for $A_-$ is
\begin{equation}
	A_-=4\theta(u)\frac{f(x^-)}{g^2}\exp\left[-\frac{1}{2}g^2|u|\right]\,,
\end{equation}
while we have set $A_+$ identically to zero because it is not normalizable being proportional to
$\exp[g^2|u|/2]$. The boundary conditions constraint also $A_-$ to be identically zero,
because only the trivial solution for $f(x^-)$ allows the field to be continuous at $u=0$.\\
The particular solution found in the aforementioned work is not present in the dynamics of the low energy theory of the configuration $D2$- $D8$-brane.

\section{Localized Chiral Fermion in $\mathbf{\mathbb{R}^{3,1}}$}
\label{codim2}
We now generalize the previous setup to a model with a four dimensional bulk and chiral fermion localized on a two dimensional topological defect. The model is composed of a $U(1)$ gauge field $A_M$ and a scalar field $C^{(0)}$ propagating in the bulk, and chiral fermions localized on the defect, topologically a string,
\begin{eqnarray}
S=\int\de^4x\left[-\frac{1}{4g^2}F_{AB}^2
	-\frac{1}{2}\epsilon^{ABCD}\partial_AC^{(0)}A_BF_{CD}+\frac{1}{2}(\partial_MC^{(0)})^2+\right.\nonumber\\
	\left.+J_{AB}H^{AB}+\delta^{(2)}\,(\vec u)
	\bar\psi_\mathrm{L}(\imm\nslash\partial+\nslash{A})\psi_\mathrm{L}\right]\,.
\end{eqnarray}
The fermion is naturally coupled to the gauge field; the tensor $J_{AB}$ is the string current defined as
$J_{AB}=\int\de^2\xi\delta(x-Y(\xi))\partial^{\,a}Y_A\partial^{\,b}Y_B\,\epsilon_{ab}$,
and $H^{AB}$ is the magnetic dual of the scalar $C^{(0)}$, $dH=\star dC^{(0)}$ where $\star$ is the Hodge star product in $\mathbb{R}^{3,1}$. This model can
be seen as the effective action for the following brane configuration in type IIB string theory:
one $D3$-brane intersecting $D7$-branes on a 1+1 worldsheet, i.e. the string. The fields we are considering are the massless open string modes of such configuration. Some dimensions on the worldvolume of the
$D$-branes are non supersymmetrically compactified and so the adjont fermions, that should appear in
the massless spectrum, are decoupled from the low energy theory.
For sake of simplicity we are neglecting the modes coming from the $D7$-branes.\\
The scalar field $C^{(0)}$ is understood as the RR field in the IIB spectrum, and its
interaction with the gauge field is part of the Chern-Simon term, formally
written as $C\wedge\exp[F]$, where $C\equiv\sum C^{(i)}$ and $C^{(i)}$ are the Ramond-Ramond fields.\\
The string is magnetically coupled to the $C^{(0)}$ form.\\
We consider cylindrically symmetric configurations and so the effective action after
integrating around the string is
\begin{equation}
\l=u\left[-\frac{1}{4g^2}F_{MN}^2\right]-\frac{1}{2}\epsilon^{MNP}A_MF_{NP}+\delta(u)\,\frac{1}{2}A_\mu^2\,,
\end{equation}
the integral of $C^{(0)}$ is fixed by the string magnetic charge.
We have implicitly integrated out the fermion field as well. On lightcone coordinates the Lagrangian is
\begin{equation}
\l=u\left[\frac{1}{2g^2}F_{+-}^2-\frac{1}{g^2}F_{u-}F_{u+}\right]-\frac{1}{2}(A_+F_{-u}+A_-F_{u+}+A_uF_{+-})
	+\delta(u)A_+A_-\,,
\end{equation}
and by differentiating with respect to the gauge field, the equations of motion are
\begin{eqnarray}
 F_{u-}+\frac{1}{g^2}(u\,\partial_-F_{+-}+\partial_uuF_{u-})+\delta(u)A_-&=&0\,,\\
-F_{u+}+\frac{1}{g^2}(u\,\partial_+F_{-+}+\partial_uuF_{u+})+\delta(u)A_+&=&0\,,\\
 F_{+-}+\frac{u}{g^2}(\partial_+F_{u-}+\partial_-F_{u+})&=&0\,.
\end{eqnarray}
Once again, a differential equation for $F_{+-}$ can be recovered after algebraic manipulations. The Kaluza Klein
condition is $2\partial_+\partial_-F_{+-}=\mu^2F_{+-}$, and so the equation is
\begin{equation}
\label{codim2.fpm}
-u\,\partial_u\,u\,\partial_u\,F_{+-}+(g^4-\mu^2u^2)F_{+-}
	+\frac{u}{g^2}\partial_u(u\,\delta(u)\partial_\mu A^\mu)-\delta(u)\,uF_{+-}=0\,.
\end{equation}
We have retained the localized term to show that the third term in the above expression can be removed
by a suitable gauge choice and is therefore a gauge artifact. Equation
\eqref{codim2.fpm} is scale invariant in the $u$ direction, hence
we can substitute $u$ with the adimensional coordinate $\u=\mu u$, and the only parametrical dependency is on the adimensional coupling constant $g$:
\begin{equation}
-\u\,\partial_{\u}\,\u\,\partial_{\u}F_{+-}+(g^4-\u^2)F_{+-}-\delta(\u)\,\u F_{+-}=0\,.
\end{equation}
In order to solve the equation and impose the boundary conditions, we regularize the $\delta$ function as
$\theta(\e-\u)/\e$, and we take the zero $\e$ limit to recover the thin wall approximation we are considering.
The equations for $F_{+-}$ in the two regions are:
\begin{eqnarray}
-\u^2\,F_{+-}^{\prime\prime}-\u F_{+-}^\prime+(g^4-\u^2)F_{+-}&=&0\quad\u>\e\,,\\
-\u^2\,F_{+-}^{\prime\prime}-\u F_{+-}^\prime+(g^4+\frac{\u}{\e}-\u^2)F_{+-}&=&0\quad\u<\e\,,
\end{eqnarray}
which can both be solved analytically. The solution for the first equation is a sum of Bessel functions of the
first and second kind, while for the second it is a sum of confluent hypergeometric functions, as shown explicitly
below
\begin{eqnarray}
F_{+-}&=&f^{(1)}J_{g^2}(\u)+f^{(2)}Y_{g^2}(\u)\,,\\
F_{+-}&=&\u^{g^2}\esp^{-\imm\u}\left[h^{(1)}M(\frac{\imm}{2\e}+\frac{1+2g^2}{2},1+2g^2,2\imm\u)+\right.\nonumber\\
      &&\phantom{\u^{g^2}\esp^{-\imm\u}}\left.+h^{(2)}U(\frac{\imm}{2\e}+\frac{1+2g^2}{2},1+2g^2,2\imm\u)\right]\,.
\end{eqnarray}
To avoid clutter, we shorten the two confluent hypergeometric functions with $\bar M(\u)$
and $\bar U(\u)$. By imposing regularity of the solution at the origin, given the string has a finite thickness,
we choose the coefficient $h^{(2)}$ to be zero because the hypergeometric function $U$ diverges for zero argument.\\
The boundary conditions we have to impose are continuity and differentiability of the solution at $\u=\e$, the
former gives
\begin{equation}
\label{codim2.bc.cont}
f^{(1)}J_{g^2}(\e)+f^{(2)}Y_{g^2}(\e)=\e^{g^2}\esp^{-\imm\e}h^{(1)}\bar M(\e)\,,
\end{equation}
and the latter
\begin{equation}
\label{codim2.bc.diff}
f^{(1)}J_{g^2}^\prime(\e)+f^{(2)}Y_{g^2}^\prime(\e)=\e^{g^2}\esp^{-\imm\e}h^{(1)}
	\left[\left(\frac{g^2}{\e}-\imm\right)\bar M(\e)+\bar M^\prime(\e)\right]\,.
\end{equation}
Solving for $h^{(1)}$ in \eqref{codim2.bc.cont} and substituting in \eqref{codim2.bc.diff},
we find the following relation between the coefficients $f^{(1)}$ and $f^{(2)}$:
\begin{equation}
\frac{f^{(2)}}{f^{(1)}}=-\frac
	{J_{g^2}-\left(\frac{g^2}{\e}-\imm+\frac{\bar M^\prime(\e)}{\bar M(\e)}\right)J_{g^2}^\prime(\e)}
	{Y_{g^2}-\left(\frac{g^2}{\e}-\imm+\frac{\bar M^\prime(\e)}{\bar M(\e)}\right)Y_{g^2}^\prime(\e)}\,.
\end{equation}
One more constraint must be imposed, as usual we have to fix the normalization of the vector field to be one.
At large values of $\u$ the Bessel functions can be approximated with decreasing plane waves
\begin{equation}
J_{g^2}(\u)\sim\sqrt\frac{2}{\pi\u}\cos\left[\u-g^2\frac{\pi}{2}\right]\,,\qquad
Y_{g^2}(\u)\sim\sqrt\frac{2}{\pi\u}\sin\left[\u-g^2\frac{\pi}{2}\right]\,.
\end{equation}
Therefore the normalization condition for $f^{(i)}$ is $(f^{(1)})^2+(f^{(2)})^2=1$. The solutions
of these two constraints, boundary conditions and normalization, are\footnote{attention should
be paid for $\bar M^\prime(\e)/\bar M(\e)$ in the limit for small $\e$ since both the argument and the
parameters depend on $\e$; we remind to the appendix \ref{app3} for a detailed discussion.}
\begin{eqnarray}
f^{(1)}&=&\phantom{-}\frac
	{Y_{g^2}(\e)-\e(g^2-\frac{1}{1+2g^2})^{-1}Y_{g^2}^\prime(\e)}
	{\sqrt{(J_{g^2}(\e)-\e\frac{1+2g^2}{2g^4+g^2-1}J_{g^2}^\prime(\e))^2+
	       (Y_{g^2}(\e)-\e\frac{1+2g^2}{2g^4+g^2-1}Y_{g^2}^\prime(\e))^2}}\,,\\
f^{(2)}&=&-\frac
	{J_{g^2}(\e)-\e(g^2-\frac{1}{1+2g^2})^{-1}J_{g^2}^\prime(\e)}
	{\sqrt{(J_{g^2}(\e)-\e\frac{1+2g^2}{2g^4+g^2-1}J_{g^2}^\prime(\e))^2+
	       (Y_{g^2}(\e)-\e\frac{1+2g^2}{2g^4+g^2-1}Y_{g^2}^\prime(\e))^2}}\,.
\end{eqnarray}
In the limit $\e\rightarrow0$
\begin{eqnarray}
f^{(1)}&\sim&1\,,\\
f^{(2)}&\sim&\frac{2\pi}{\Gamma[g^2]\Gamma[1+g^2]}\frac{4g^2-3}{4g^4+4g^2-1}\left(\frac{\e}{2}\right)^{2g^2}
	\rightarrow0\,.
\end{eqnarray}
The solution for the stress tensor field is
\begin{eqnarray}
F_{+-}&=&\phantom{-}f(x^\pm)J_{g^2}(\mu u)\,,\\
F_{u+}&=&         - \frac{\partial_+f}{\mu}J_{g^2-1}(\mu u)\,,\\
F_{u-}&=&         - \frac{\partial_-f}{\mu}J_{g^2+1}(\mu u)\,.
\end{eqnarray}
As we have already seen for the case of $D2$-brane intersecting $D8$-brane, also for this case, the
solution has no support at the origin. The electromagnetic field is delocalized completely in the bulk and
is not allowed to penetrate the interior of the string.\\
This result seems to be independent of the codimensions and therefore we propose that it is still present
for greater codimensions. Unfortunately no analytical solutions is known for codimensions $n>2$.

\paragraph{Solution with $\mathbf{F_{+-}=0}$.}
Following the same scheme as before, we may now want to look for solution of the kind $F_{+-}=0$. The equations of motion with
gauge choice $A_u=0$ are:
\begin{eqnarray}
\partial_u\,u\,\partial_u\,A_-+g^2\partial_u\,A_-+\delta(u)\,g^2A_-&=&0\,,\\
\partial_u\,u\,\partial_u\,A_+-g^2\partial_u\,A_++\delta(u)\,g^2A_+&=&0\,,\\
\partial_u(\partial_+A_-+\partial_-A_+)&=&0\,.
\end{eqnarray}
The last equation with $F_{+-}=0$ imposes the vector field $A_\pm$ to be holomorphic in $x^\pm$. The solutions are
\begin{equation}
A_\pm=\pm\frac{f_\pm}{g^2}u^{\pm g^2}\,,
\end{equation}
and the coefficient $f_+$ for $A_+$ has to be zero since this component is not normalizable.\\
We are now left with finding the coefficient $f_-$ by imposing the boundary conditions. Following closely
the technique used, before the equations of motion with regularized $\delta$ function are
\begin{eqnarray}
g^2\partial_u\,A_-+\partial_u\,u\,\partial_u\,A_-&=&0\quad\,u>\e\,,\\
g^2\partial_u\,A_-+\partial_u\,u\,\partial_u\,A_-+\frac{1}{\e}A-&=&0\quad\,u<\e\,,
\end{eqnarray}
with the following regular solutions:
\begin{eqnarray}
&A_-=-\frac{f_-}{g^2}u^{-g^2}&\quad\,u>\e\,,\\
&A_-=\phantom{-}h\,\Gamma[1+g^2]J_{g^2}\left[2g\sqrt{u/\e}\right]
	\left(\frac{\e}{ug^2}\right)^{g^2/2}&\quad\,u<\e\,.
\end{eqnarray}
The boundary conditions impose two constraints on $f_-$
\begin{eqnarray}
f_-&=&-h\,\Gamma[1+g^2]\,g^{2-g^2}\e^{g^2}J_{g^2}(2g)\,,\\
f_-&=&-h\,g^{2-g^2}\e^{g^2}J_{1+g^2}(2g)\,,
\end{eqnarray}
which translate into the following relation:
\begin{equation}
J_{1+g^2}(2g)=\Gamma[1+g^2]J_{g^2}(2g)\,.
\end{equation}
Since there is no value of $g$ that satisfies this relation, as before we find no solution of this kind.

\section{Generic Dimension Bulk}
\label{codimgen}

As we said already, no exact solutions are known for codimensions greater than two, but we can still look for
a solution, that satisfies $F_{+-}=0$. Taking $A_u=0$, the equations of motions for generic $n>2$ are
\begin{eqnarray}
\partial_uu^{n-1}\partial_u\,A_+-g^2\partial_u\,A_++\delta(u)\,g^2A_+&=&0\,,\\
\partial_uu^{n-1}\partial_u\,A_-+g^2\partial_u\,A_-+\delta(u)\,g^2A_-&=&0\,,\\
\partial_u(\partial_+A_-+\partial_-A_+)&=&0\,,
\end{eqnarray}
with the following solutions
\begin{equation}
A_\pm=\pm\frac{f_\pm}{g^2}\left\{\exp\left[\pm\frac{g^2}{n-2}\frac{1}{u^{n-2}}\right]-1\right\}\,,
\end{equation}
where $f_\pm$ are coefficient functions holomorphic in $x^\pm$ to be fixed by the boundary conditions on the
string. We have chosen the integration constant in such a way that the vector field is null at infinity.\\
Since $u^{n-1}|A_-|^2\sim u^{3-n}$ when $u\rightarrow\infty$, only for $n\ge5$ the vector field is
normalizable. The coefficient for $A_+$ is set to zero, for this component strongly diverges at $u\rightarrow0$.\\
For any value of $n>2$, $1/A_-$ has an essential singularity at $u=0$ and therefore the boundary condition
just imposes $A_-(0)\sim f_-(x^-)$ to be zero.\\
We conclude that for any codimension greater than two - or for any
codimension at all remembering the previous results - such a solution does not exist.

\section{Discussion}
\label{disc}

In this paper we have shown that the anomaly inflow forces the bulk gauge field away from the defects where chiral fermions are localized. From a low energy perspective, this turns into the decoupling of the zero-mode from the two dimensional spectrum. We discussed how this dynamics is equivalent to the Coleman's theorem for two dimensional field theories.\\
We analytically solved the dynamics of the gauge field for two solvable brane models with two dimensional intersection. The first was constructed from one $D2$-brane intersecting $D8$-branes. Its low energy spectrum consists of (2+1)-dimension gauge field and chiral fermions localized on a (1+1)-dimension defect. The non-zero anomaly is cancelled by anomaly inflow from the bulk. We have seen that the anomaly cancellation mechanism effectively induces a repulsive potential for the gauge field in the equivalent Schr\"odinger problem. The effect is to force the field away from the intersection and thus removing its zero-mode from the low energy spectrum.\\
We have also solved the analogous dynamics for a brane configuration consisting of one $D3$-brane intersecting $D7$-branes. The dynamics is analogous to the one discussed previously. The anomaly inflow induces a repulsive potential, that effectively expels the gauge field away from the intersection. Because the gauge field has no support on the intersection, its zero-mode is decoupled from the low energy two dimensional spectrum.\\
From a low energy perspective, the zero-mode of the gauge field describes the Nambu-Goldstone boson associated with a spontaneously broken symmetry. Its absence is expected from Coleman's theorem, and the described dynamics enforces the theorem in the dual picture.

The generalization to higher dimensions has been found hard to solve for the non-existence of an analytic solution for the generalized equation of motion for $F_{+-}$, as in \eqref{eqF+-} and \eqref{codim2.fpm},
\begin{equation}
-u^{n-1}\partial_u\,u^{n-1}\,\partial_u\,F_{+-}+(g^4-\mu^2u^{2n-2})F_{+-}=0\,,
\end{equation}
where $n$ is the codimension of the intersection ($n=1$ in section~\ref{codim1}, and $n=2$ in section~\ref{codim2}.) In section~\ref{codimgen}, we have discussed a particular class of solutions and found that they are not present in the spectrum. It would be interesting to be able to solve at least numerically the higher dimensional case and to show the dynamics for cases such as the one of \cite{Antonyan:2006qy} or \cite{Itzhaki:2005tu}. Further investigation may be required in this direction.

It would be interesting to study in greater details what are the effects of the delocalization on closed string modes. In \cite{Marolf:1999uq} and \cite{Gomberoff:1999ps}, it was shown that in certain cases when a $D$-brane is fully contained into another, the former delocalizes in the worldvolume of the latter. Since those configurations are $T$-dual to the ones we studied, a similar effect may be expected. The details, however, seem to be far less clear.\footnote{I would like to thank Don Marolf for pointing me to those works.}

Another generalization of our discussion would be to study the dynamics for non-abelian groups, that to say for $N_f$ branes intersecting with $N_c$ branes. The noncommutativity of the gauge field adds technical difficulties for solving analytically the equations of motion and finding the Kaluza-Klein spectrum.\\
The generalization of the discussion can be easily implemented. For instance, in section~\ref{codim2}, we can consider $N_f$ coincident $D3$-branes intersecting with $D7$-branes, instead of only one $D3$-brane. As mentioned previously, chiral fermions are localized on the 1+1 topological defect, and the RR scalar $C^{(0)}$ and the $SU(N_f)$ non-abelian gauge field $A_M$ are propagating in the bulk. The interactions are between fermions and gauge field on the string, and between RR scalar and gauge field in the bulk. The latter has the form
\begin{equation}
\int dC^{(0)}\wedge\omega_3\quad\mbox{where}\quad
\omega_3\equiv\mathrm{Tr}[\,A\wedge F+\frac{2}{3}A\wedge A\wedge A\,]\,.
\end{equation}
$\omega_3$ is the Chern-Simon form for non-abelian groups and generalizes the abelian case $\omega_3\equiv A\wedge F$. The presence of the Chern-Simon term is essential for the consistency of the model, since otherwise the gauge current would be non-conserved because of chiral anomaly.\\
The intersection is magnetically charged under the scalar field $C^{(0)}$, which can be integrated out and whose flux is set by the charge on the string. An effective three dimensional interaction, analogous to the one discussed in section~\ref{codim2}, is recovered. We suggest, because of the analogy with the already studied cases, that the effective Chern-Simon term generates a repulsive potential for the equivalent Schr\"odinger equation and forces the gauge field to delocalize in the bulk: no zero-mode is present in the spectrum and this ``disappearance" explains the absence of the NG boson as required by Coleman's theorem.

\section*{Acknowledgment}
I would like to thank G.~Gabadadze for deep insights and unrelenting patience. I am grateful to O.~Aharony, F.~Nitti and O.~Pujolas for enlightening conversations, and S.~Pasquali, M.~Brigante and L.~Huang for helpful discussions.
Research supported by Graduate Students funds provided by New York University.

\section*{Appendix}
\appendix
\section{Bosonization of Dirac Schwinger model}
\label{app1}

In 1+1 dimensions, fermions are known to be equivalent to bosons; a map can be constructed that links a
fermion field $\psi$ to a boson field $\phi$. A set of rules can be provided to equate the fermion
currents, both vector $\bar\psi\gamma^\mu\psi$ and axial $\bar\psi\gamma^\mu\gamma^5\psi$, with their
correspondent bosonic parts. Given these rules the boson Lagrangian is built, and its dynamics is equivalent to the one for the fermion Lagrangian.\\
In the present appendix we review the bosonization of the Dirac Schwinger model, namely the model for Dirac
fermions coupled to a $U(1)$ gauge field in 1+1 dimensions; in the next appendix we will discuss
the same construction for chiral fermions.\\
The Lagrangian is
\begin{equation}
S=\int\de^2x\,\left[-\frac{1}{4}F_{\mu\nu}^{\,2}+\bar\psi(\imm\nslash\partial+\nslash A)\psi\right]\,,
\end{equation}
where we set the charge to 1 for sake of clarity. The map between the fermion $\psi$ and its
equivalent boson $\phi$ is given by $\psi\sim\exp[\imm\gamma^5\phi]$. Given this equivalence, the
bosonization rules are:
\begin{eqnarray}
\bar\psi\imm\nslash\partial\psi&=&\frac{1}{2}(\partial_\mu\,\phi)^2,\label{dbosrule1}\\
\bar\psi\gamma^\mu\psi&=&\epsilon^{\mu\nu}\partial_\nu\phi\,,\label{dbosrule2}\\
\bar\psi\gamma^\mu\gamma^5\psi&=&\eta^{\mu\nu}\partial_\nu\phi\,.\label{dbosrule3}
\end{eqnarray}
The previous relations may be found by taking the two fermion fields in each operator at slightly
different points and expand the exponential over the difference $\varepsilon$ in the position of
$\bar\psi$ and $\psi$. Taking the leading order, and renormalizing the divergence for $\varepsilon\to0$,
leads to the previous set of rules.\\
We can now substitute \eqref{dbosrule1}, \eqref{dbosrule2} and \eqref{dbosrule3} into the fermion action
\begin{equation}
S=\int\de^2x\,\left[-\frac{1}{4}F_{\mu\nu}^{\,2}+\frac{1}{2}(\partial_\mu\,\phi)^2
	+\epsilon^{\mu\nu}A_\mu\partial_\nu\phi\right]\,.
\end{equation}
The equivalent bosonized model is of a massless scalar field $\phi$ coupled to the gauge field $A_\mu$.
The coupling is the reminiscence of the chiral anomaly in the fermion action. Given the exponential
representation of the field $\psi$ in terms of $\phi$, it is straightforward to see that a chiral rotation
of $\psi$ corresponds to a shift of $\phi$; by acting with this transformation on the action we find
$\delta S=\epsilon^{\mu\nu}\partial_\mu A_\nu$, which is the chiral anomaly in two dimensions.

\section{Bosonization of Chiral Schwinger model}
\label{app2}

In the present appendix we discuss the previous technique for bosonizing a fermion field in the case
of the chiral Schwinger model. The spectrum comprises a $U(1)$ gauge field $A_\mu$ and a chiral fermion
$\psi_\mathrm{L}\sim(1-\gamma^5)\psi$. For these fields, the action is:
\begin{equation}
S=\int\de^2x\,\left[-\frac{1}{4}F_{\mu\nu}^{\,2}
	+\bar\psi_\mathrm{L}(\imm\nslash\partial+\nslash A)\psi_\mathrm{L}\right]\,.
\end{equation}
The bosonization rules resemble the ones given before for the non-chiral case,
\begin{eqnarray}
\psi_\mathrm{L}\imm\nslash\partial\psi_\mathrm{L}&=&\frac{1}{2}(\partial_\mu\,\phi)^2\,,\label{bosrule1}\\
\bar\psi_\mathrm{L}\gamma^\mu\psi_\mathrm{L}&=&(\eta^{\mu\nu}-
	\epsilon^{\mu\nu})\partial_\nu\phi+\frac{1}{2}a A^\mu\,,\label{bosrule2}
\end{eqnarray}
the main difference is for the fermion current \eqref{bosrule2}. The first term can be
understood by considering that the current is proportional to the sum of the vector and the axial
current, and so using the previous rules \eqref{dbosrule2}, \eqref{dbosrule3} we are lead to it. The
term proportional to the gauge field has no direct explanation at this point, but we may point out that
it can always be added. In the previous case, the addition of $aA_\mu$ is constrained by the
requirement of gauge invariance and therefore $a$ is set to zero. The chiral action is not invariant
under gauge transformation $\delta A_\mu=\partial_\mu\Lambda$ because of the anomaly, therefore the parameter
can not be fixed at the moment; we will see that the requirement of a ``minimal" anomaly will fix $a$.\\
We may want to write the bosonization rules for the vector and the axial currents as
\begin{eqnarray}
\bar\psi\gamma^\mu\psi&=&\epsilon^{\mu\nu}\partial_\nu\phi+\epsilon^{\mu\nu}A_\nu\,,\label{bosrule1-eq}\\
\bar\psi\gamma^\mu\gamma^5\psi&=&\eta^{\mu\nu}\partial_\nu\phi-\epsilon^{\mu\nu}A_\nu+a\eta^{\mu\nu}A_\nu\,.\label{bosrule2-eq}
\end{eqnarray}
The relations \eqref{bosrule1} and \eqref{bosrule2} are found by summing up the rules \eqref{bosrule1-eq} and \eqref{bosrule2-eq}.\\
The presence of the added term $\epsilon^{\mu\nu}A_\nu$ is to ensure the vector current to be invariant under gauge transformation, $\delta A_\mu=\partial_\mu\Lambda$ and $\delta\phi=-\Lambda$.
The field $\phi$ transforms under gauge transformation, because the fermion field $\psi$
is charged; since $\psi_\mathrm{L}=\exp[\imm\phi]$, $\delta\psi_\mathrm{L}=\exp[-\imm\Lambda]$ imposes the field
$\phi$ to shift under gauge transformation. Therefore variation of the vector current is $\delta j^\mu=0$, instead
the axial current transforms as
$\delta j_\mu^5=(a-1)\eta_{\mu\nu}\partial^\nu\Lambda-\epsilon_{\mu\nu}\partial^\nu\Lambda$. The result is
understood since the axial current is not conserved because of quantum corrections, which appear at classical level
in the bosonized model.\\
Substituting \eqref{bosrule1} and \eqref{bosrule2} into the action, we obtain
\begin{equation}
S=\int\de^2x\,\left[-\frac{1}{4}F_{\mu\nu}^{\,2}+\frac{1}{2}(\partial_\mu\,\phi)^2+A_\nu(\eta^{\mu\nu}
	-\epsilon^{\mu\nu})\partial_\mu\phi+\frac{1}{2}aA_\mu^2\right]\,,
\end{equation}
namely the action for a massless scalar field $\phi$ coupled to a gauge field $A_\mu$. In the bosonized theory
the gauge field is not massless anymore, but it acquires a mass proportional to the parameter $a$, which we shall fix.\\
Because of the anomaly, the action is not invariant under gauge transformation
\begin{equation}
\delta S=-\int\de^2x\,\left[(a-1)\partial^\mu A_\mu+\epsilon^{\mu\nu}\partial_\mu A_\nu\right]\Lambda\,,
\end{equation}
in the brackets we find the anomaly $\mathcal{A}=(a-1)\partial^\mu A_\mu+\epsilon^{\mu\nu}\partial_\mu A_\nu$.
We can see now that the requirement of a ``minimal" anomaly, namely the anomaly to be
$\mathcal{A}=\epsilon^{\mu\nu}\partial_\mu A_\nu$, fixes the choice of the parameter $a$ to be 1.

\section{Properties of Confluent Hypergeometric Functions}
\label{app3}
In this appendix we justify the asymptotic expansion for small $\e$ of the ratio between $\bar M(\e)$ and
$\bar M'(\e)$ used in section~\ref{codim2} while studying the boundary conditions for $F_{+-}$. The functions
$\bar M$ and $\bar M'$ were defined as
\begin{eqnarray}
\bar M(\e)&=&M(\frac{\imm}{2\e}+\frac{1+2g^2}{2},1+2g^2,2\imm\e)\,,\\
\bar M'(\e)&=&M'(\frac{\imm}{2\e}+\frac{1+2g^2}{2},1+2g^2,2\imm\e)\,,
\end{eqnarray}
where $M(a,b,z)$ is the confluent hypergeometric function of the first kind. The above functions
dependend on $\e$ both in the argument and in the parameters, and one could question the effective convergence
of $\bar M(\e)/\bar M'(\e)$.\\
This particular hypergeometric function admits a series expansion of the form:
\begin{equation}
\bar M(\e)=\sum_{k=0}^\infty\frac{\left(\frac{\imm}{2\e}+\frac{1+2g^2}{2}\right)_k}{(1+2g^2)_k}\,
	\frac{(2\imm\e)^k}{k!}\,,
\end{equation}
and
\begin{equation}
\bar M'(\e)=\left[-\frac{1}{(1+2g^2)\e}+\imm\right]
	\sum_{k=0}^\infty\frac{\left(\frac{\imm}{2\e}+\frac{3+2g^2}{2}\right)_k}{(2+2g^2)_k}\,
	\frac{(2\imm\e)^k}{k!}\,.
\end{equation}
The symbol $(x)_k$ is the Pochhammer symbol defined as $(x)_k\equiv \Gamma[x+k]/\Gamma[x]$ and is equal to one for
any non-negative integer argument. Because of this last property, we can easily take the first order approximation in $\e$
for the ratio $\bar M'(\e)/\bar M(\e)$ and find
\begin{equation}
\frac{\bar M'(\e)}{\bar M(\e)}\sim-\frac{1}{(1+2g^2)\e}+\imm\,.
\end{equation}
\bibliographystyle{utphys}
\bibliography{chernsimon}
\end{document}